# A low-temperature ultra-high-vacuum scanning probe microscope with in situ electronic transport capabilities: from macro to nano in 10 minutes

Diego Expósito[1,2,3], Oscar Custance[4], Iván Brihuega[1,2,3]

1 Departamento de Física de la Materia Condensada, Universidad Autónoma de Madrid, E-28049 Madrid, Spain

2 Condensed Matter Physics Center (IFIMAC), Universidad Autónoma de Madrid, E-28049 Madrid, Spain

3 Instituto Nicolás Cabrera (INC), Universidad Autónoma de Madrid, E-28049 Madrid, Spain

4 National Institute for Materials Science (NIMS), 305-0047 Tsukuba, Ibaraki, Japan.

We have developed a low temperature (LT), ultra-high-vacuum (UHV) system that combines two complementary techniques, scanning probe microscopy (SPM) and electrical transport measurements, within a single platform. By providing simultaneous access to the atomic-scale surface landscape and the macroscopic device response of the same sample, the setup enables direct correlations between local structural/spectroscopic signatures and global electronic transport behavior in two-dimensional (2D) devices. The system allows experiments where atomic-scale modifications or controlled manipulations are performed while continuously monitoring their impact on device-scale performance. The setup consists of two interconnected UHV chambers: a dedicated preparation chamber and a separate measurement chamber that houses a liquid-helium cryostat and the SPM/transport stage. Base pressure is $1\times10^{-11}$ Torr. A key feature is direct optical access to the sample, enabling rapid and reliable tip positioning with an accuracy of $5 \times 5\ \mu m^2$ within 10 minutes. The SPM, operated using custom-built electronics, can track the exact same sample region across a temperature range from 2.9 K to 400 K, with mechanical stability below 1 pm. System performance is demonstrated on graphene devices, and bulk Pb is used to determine energy resolution. Using superconducting tips, scanning tunneling spectroscopy measures the superconducting gap with an energy resolution of 30 μV. Transport measurements track the temperature dependence of both resistivity and critical current across the superconducting transition of an in-situ prepared Pb nanowire.

## I. Introduction

The discovery of graphene through mechanical exfoliation in 2004 gave birth to an exciting field of research focused on the exploration of 2D materials, which remains at the forefront of materials science and condensed matter research [1–7]. Progress in this field has greatly benefited from the use of complementary characterization tools. Quantum transport measurements and scanning tunneling microscopy (STM) have been two widely employed techniques to study the electronic properties of 2D materials and vdW heterostructures. Each of these techniques has its own set of strengths. On the one hand, transport experiments enable the study of the global properties of 2D materials. Using multiple gates, one can tune electrostatic potentials to control, e.g., the charge carrier density and the electric displacement field. On the other hand, STM constitutes a powerful tool to obtain local information with atomic resolution. Both topography and electronic properties are routinely mapped with atomic spatial and high energy resolution. In addition, STM manipulation capabilities enable the arrangement, with atomic precision, of atoms, molecules, or metallic species previously deposited on the surface by UHV in situ preparation machinery. This is a fundamental step, since the unique electronic properties of 2D devices are exceptionally sensitive to their environment, a characteristic that is both a challenge for device integrity and an opportunity for intentional functionalization.

While the above techniques have proven to be extremely powerful for studying 2D systems on their own, very interesting research avenues emerge when considering the possibility of combining them to probe the properties of the same sample at both global and local scales. This would enable the correlation of atomic-scale morphology with macroscopic transport response. Moreover, 2D samples could be manipulated with atomic precision by an STM and subsequently investigated by means of transport. This complementary strategy holds strong potential to uncover new physics and to clarify how local structure, disorder, and functionalization govern the electronic behavior of 2D systems. Despite these exciting prospects, integrating all these capabilities: LT, UHV sample preparation techniques, precision tip-landing, and SPM and 4-probe transport characterization within a single platform, still remains a formidable technical challenge.

Here, we present the design and performance of a new versatile LT-UHV STM/AFM system that integrates all the above-mentioned capabilities. This system is specifically optimized for the study of 2D devices based on van der Waals heterostructures by means of simultaneous STM and transport measurements, and it is equipped with a UHV preparation chamber that enables the tuning of 2D device properties through ultra-clean sample-preparation techniques. Clear optical access ensures rapid sample transfer and reliable tip landing on a micrometer-sized device within 10 minutes.

## II. Sample compatibility

One of the main challenges when integrating techniques as diverse as UHV-SPM and transport measurements is combining the different samples employed for both types of experiments. On one hand, UHV-SPM samples typically consist of few millimeter samples prepared in situ in UHV conditions, which exhibit spatially homogeneous, atomically clean surfaces. For STM in particular just one electrical contact, the tunneling bias voltage ($V_b$), is commonly needed to perform the measurements. On the other hand, 2D transport samples typically consist of μm-sized heterostructures assembled by stacking different 2D materials. Carrying out transport experiments on such devices requires a specific electrical contact configuration with at least five connections: four to perform 4-probe measurements and one for the gate voltage.

Fabricating 2D devices compatible with both techniques requires integrating polymer-based stacking methods with UHV procedures to avoid the residues typically left by the polymers used in heterostructure assembly and achieve atomically clean surfaces. While the specific methodology for achieving this level of atomic cleanliness is beyond the scope of this work, we have developed a high-precision transfer station specifically designed for this purpose [8]. This setup, inspired by the system used by the Jarillo-Herrero group at MIT [9], allows the precise sample assembly of 2D devices with a high degree of control of the stacking angle, and the obtention of atomically clean surfaces. In particular, the devices shown here consist of monolayer graphene (MLG) flakes on hexagonal boron nitride (hBN), transferred onto a $SiO_2$/Si substrate (Fig. 1c,d). External electrostatic gating is applied via a back gate using a HOPG finger or a doped $SiO_2$/Si substrate.

To simultaneously perform STM and transport experiments, we have designed a custom sample-holder based on the standard flag-style design, fully compatible with typical commercial 4-probe transport SPM microscopes (Fig. 1). The $SiO_2$/Si substrate containing the graphene device, together with the corresponding electrical pads, is mounted between two Shapal[10] ceramics. The sample-holder features six independent electrical contacts: Two tantalum plates, located on each side, provide the tunneling bias ($V_b$ ,Fig. 1a, green) and the gating voltage ($V_g$ ,Fig. 1a, red), while four spring-loaded pins are used for transport measurements (Fig. 1a blue). All connections between the sample and the sample-holder contacts are made via wire bonding. This configuration enables to carry out STM and transport measurements on the same sample while tuning its electronic properties with an external electrostatic gate voltage.

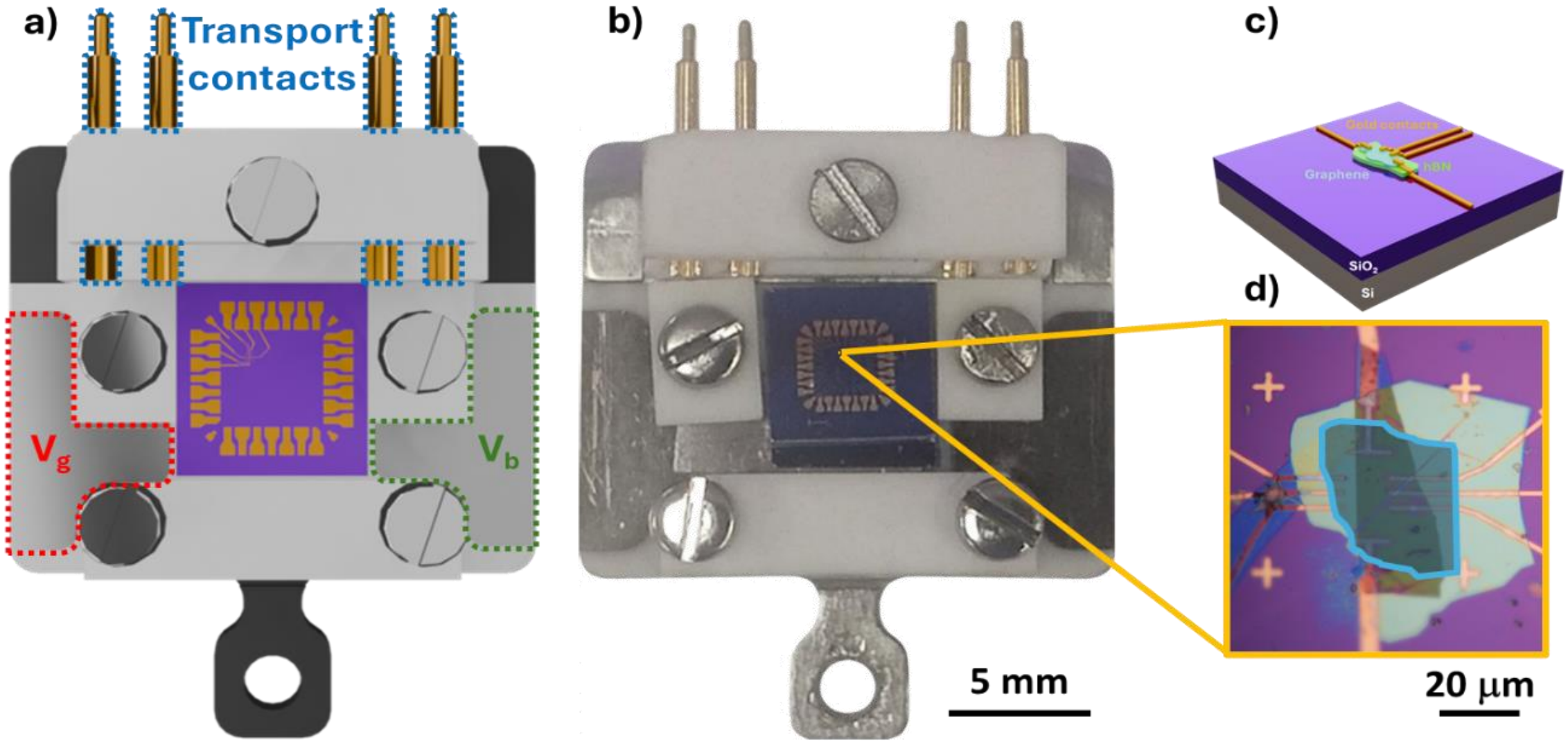

**FIG. 1.** a) 3D model of a modded flag-style sampleholder specifically designed to simultaneously perform STM and transport measurements. It has six independent electrical contacts: tunneling bias voltage (green), gate voltage (red) and four probe transport connections (blue). b) Picture of the actual sample holder modelled in a). c) 3D representation of a graphene device consisting of a MLG (blue) con top on an hBN (green) deposited on a doped $SiO_2$/Si chip. Gold contacts are deposited on top. d) Picture of an actual graphene device.

## III. UHV Experimental System

Once ultra-clean 2D devices are prepared, we introduce them into the UHV system without breaking vacuum using a load-lock chamber. Upon an initial characterization of the device, when required, we carry out a preparation in UHV to modify the device properties. To this end, we have developed a complete LT-UHV experimental platform, representing the natural evolution of previously developed home-built systems in our group (Fig. 2) [11,12]. The setup comprises two connected, custom-designed UHV chambers [13]: one dedicated to STM/transport measurements equipped with a bath cryostat [14] (Fig.2, blue), and other for sample preparation with a load-lock chamber that enables rapid introduction of new samples and probes. Typical operating conditions include a base pressure $P < 10^{-11}$ Torr and a base temperature $T = 4.2$ K, that becomes 2.9 when pumping out the helium cryostat. The entire system is mounted on four pneumatic isolators to minimize mechanical vibrations [15]. The microscope UHV chamber (Fig. 2, green) is based on the design developed by M. M. Ugeda in his PhD thesis [12], and has been adapted to enable in situ cold evaporation and improved optical access. This enhanced optical access is essential for reliably locating our graphene devices.

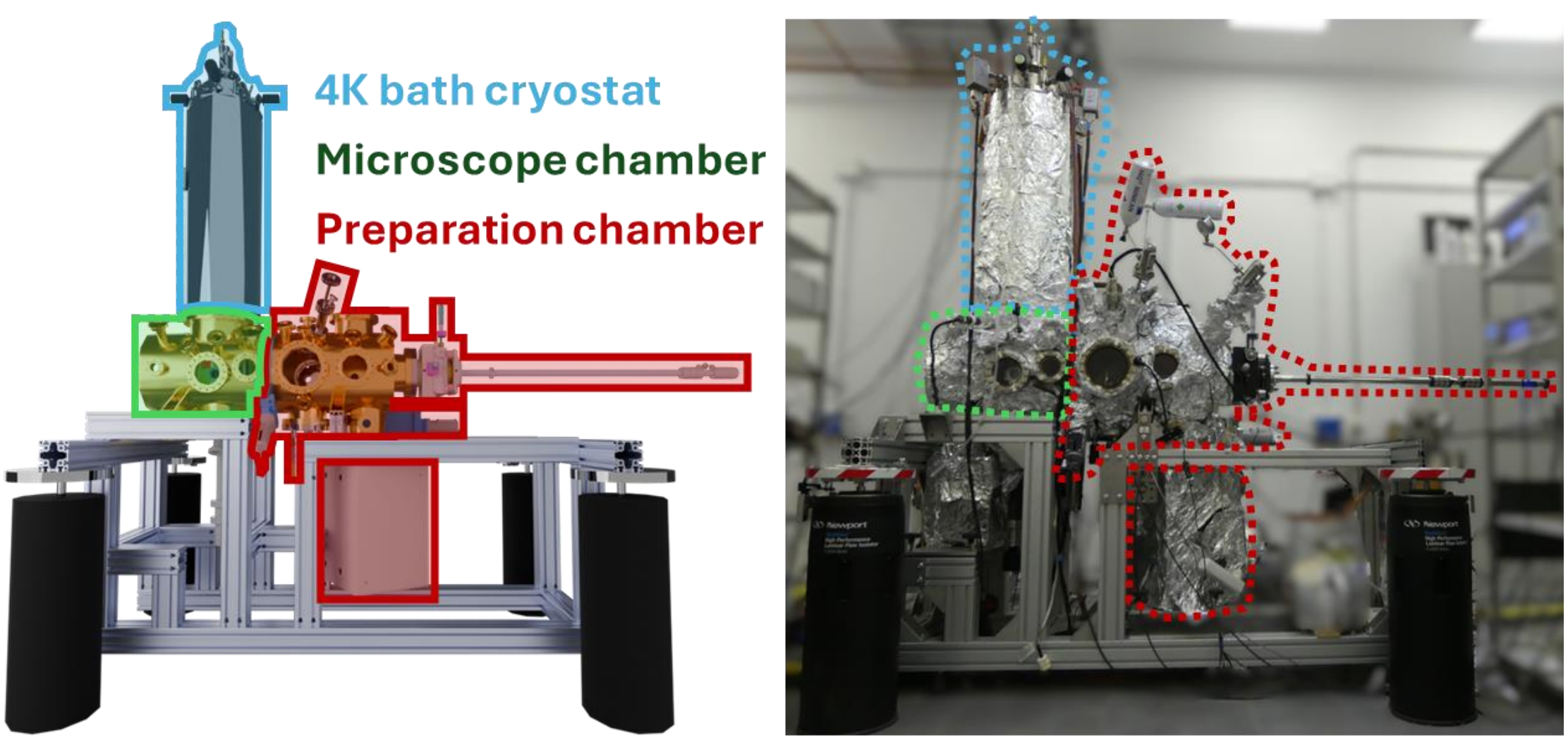

**FIG.2.** Full LT-UHV experimental system. a) Original 3D design. b) Actual picture of the setup.

The preparation chamber (Fig. 2, red) is based on the design from the PhD thesis of O. Custance[11] and incorporates the key elements required for UHV sample preparation. Samples are transferred across the system using an adapted clamp compatible with flag-style sample holders. The chamber is designed to achieve atomically clean surfaces and to enable controlled modification of the sample by depositing selected species. The chamber includes an electron-bombardment heater for annealing (Fig. 3, orange), that we mainly use to remove air adsorbates, with the option of adding direct-current heating. The sample temperature is monitored using both an optical pyrometer and a thermocouple.

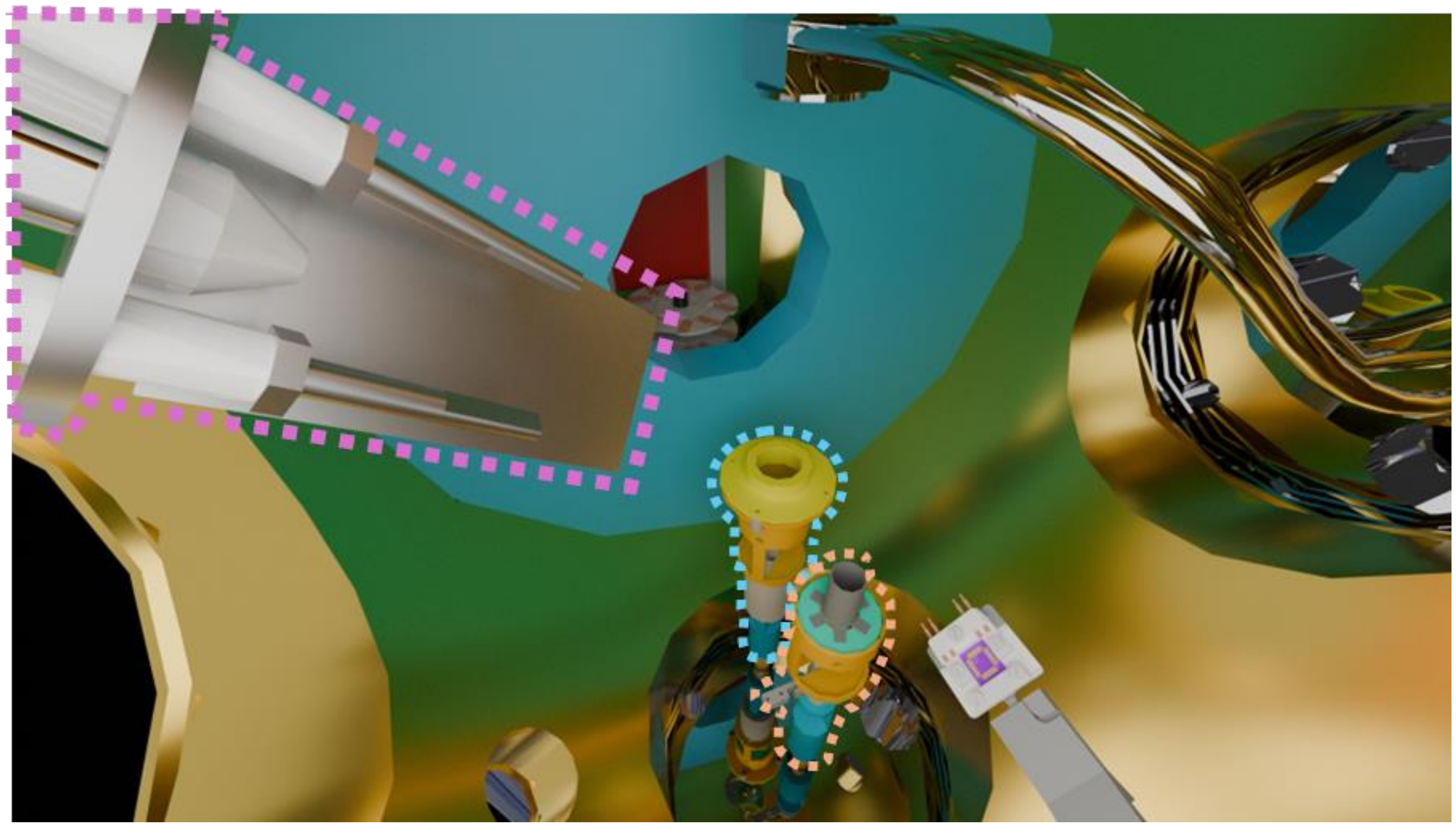

**FIG. 3.** 3D design of the preparation chamber interior. Among other components, it includes a heater (orange contour), a metal evaporator (blue contour) and a hydrogen cracker (pink contour).

Several deposition tools can be used to functionalize clean 2D-devices and samples. The chamber is equipped with an electron-beam metal evaporator with interchangeable cells, allowing different materials to be evaporated without breaking UHV (Fig. 3, blue). The evaporation rate is monitored with a quartz crystal microbalance [16]. A homemade cracker enables the delivery of atomic hydrogen, and also permits metal evaporation via resistive heating of a filament (Fig. 3, pink). An ion gun enables controlled irradiation with high-energy gas ions (e.g., $Ar^+$) to create vacancies [17], and surface cleaning by sputter-annealing cycles. Together, these tools support complex preparation protocols while maintaining the atomic cleanliness required for our measurements, and have enabled us, for example, to induce superconductivity and magnetism in graphene devices by depositing Pb and by dosing with atomic H, respectively [18,19].

## IV. STM/AFM Microscope

The microscope (Fig. 4) is mounted on a gold-plated OFHC copper base attached to the bottom of the bath cryostat. This base incorporates a clamping mechanism for safe sample and probe transfer [12]. During measurements, the microscope is freed, becoming suspended from three springs that serve as a vibration-isolation system, while a set of magnets at the bottom provides a low frequency eddy-current damping of system. To facilitate the identification and tip-landing on selected regions of 2D-devices, the microscope has a direct line of sight to the sample from outside of the UHV chamber at an angle of approximately 25°. To ensure thermal stability, wiring is thermalized through multiple stages from room temperature (RT) to 4K, effectively minimizing heat losses before reaching the microscope.

The microscope was designed with an emphasis on machining simplicity and high operational reliability. Its modular architecture allows each component to be readily attached or removed, so that repairs or modifications can be carried out straightforwardly, even with the system fully assembled. Materials were selected for their machinability and full UHV–LT compatibility. In particular, their thermal expansion coefficients are closely matched to minimize mechanical stress and prevent mechanical failure during temperature cycling from 3 K to 400 K. They also exhibit low vapor pressures to limit outgassing. To meet these requirements, the microscope is fabricated from titanium, sapphire, Shapal, gold, copper, and copper–beryllium (CuBe).

Efficient thermalization is critical for operation across such a wide temperature range. Accordingly, and also due to its relatively low density, titanium was chosen for the main body and metallic structural components. Sapphire is used for mobile parts; its extreme hardness and low friction coefficient enable smooth and precise motion of components (e.g., shear piezos) on its surface. For electrical isolation, we employ Shapal [10] ceramics, which combine high thermal conductivity with good machinability. Electrical connections are made using spring-loaded gold and CuBe pins, selected for their excellent electrical and thermal conductivities. CuBe is also used for the contact strips, as its spring constant remains nearly unchanged between 3 K and 400 K, ensuring reliable mechanical performance throughout the full temperature range.

Controlled heating is achieved using a Zener diode and two resistive heaters located inside the microscope's body [20,21]. Temperature is monitored over the full operation range with three calibrated Si diodes [22]: one at the cold finger, one close to the sample, and another close to the tip.

The microscope mainly consists of two modular stages: a tip stage (Fig. 4c) and a sample stage (Fig. 4d). The tip stage is based on the Pan design [23,24] with six shear piezo actuators for the coarse approach of the tip towards the sample. Fine atomic resolution scanning is performed with a piezoelectric tube providing a scan range of 2.2x2.2 $\mu m^2$ at 4 K. The tip stage also includes four independent electrical contacts and a tip-exchange mechanism mounted at the end of the piezoelectric tube. This enables rapid tip replacement and the use of sharp probes for reliable landing on selected regions of our graphene devices. The four independent contacts also support the integration of sensors beyond standard STM tips, such as a KolibriSensor [25] for simultaneous STM/AFM measurements. Overall, the modular design of this stage makes it readily adaptable to any compatible probe or sensor.

The sample stage provides six independent electrical contacts and an XY positioning system. The contacts are aligned with those on the sample holder, enabling simultaneous STM and transport measurements. The XY stage integrates two inertial positioners [26], offering a travel range of ±2.5 mm along both axes. Combined with excellent optical access, this capability allows precise targeting and tip landing on selected regions of our graphene devices with 5 µm accuracy, which is essential for the experiments performed on 2D devices.

The large-range XY positioning stage provides an effective way to compensate for thermal drift over the full operating temperature range of the microscope. With a lateral travel range of 5 × 5 $mm^2$, the system is designed to recover and track the same sample region while changing temperature, even when the thermally induced drift exceeds the scan range of the piezoelectric scanner [27], including during temperature changes from 40 K to 200 K [28].

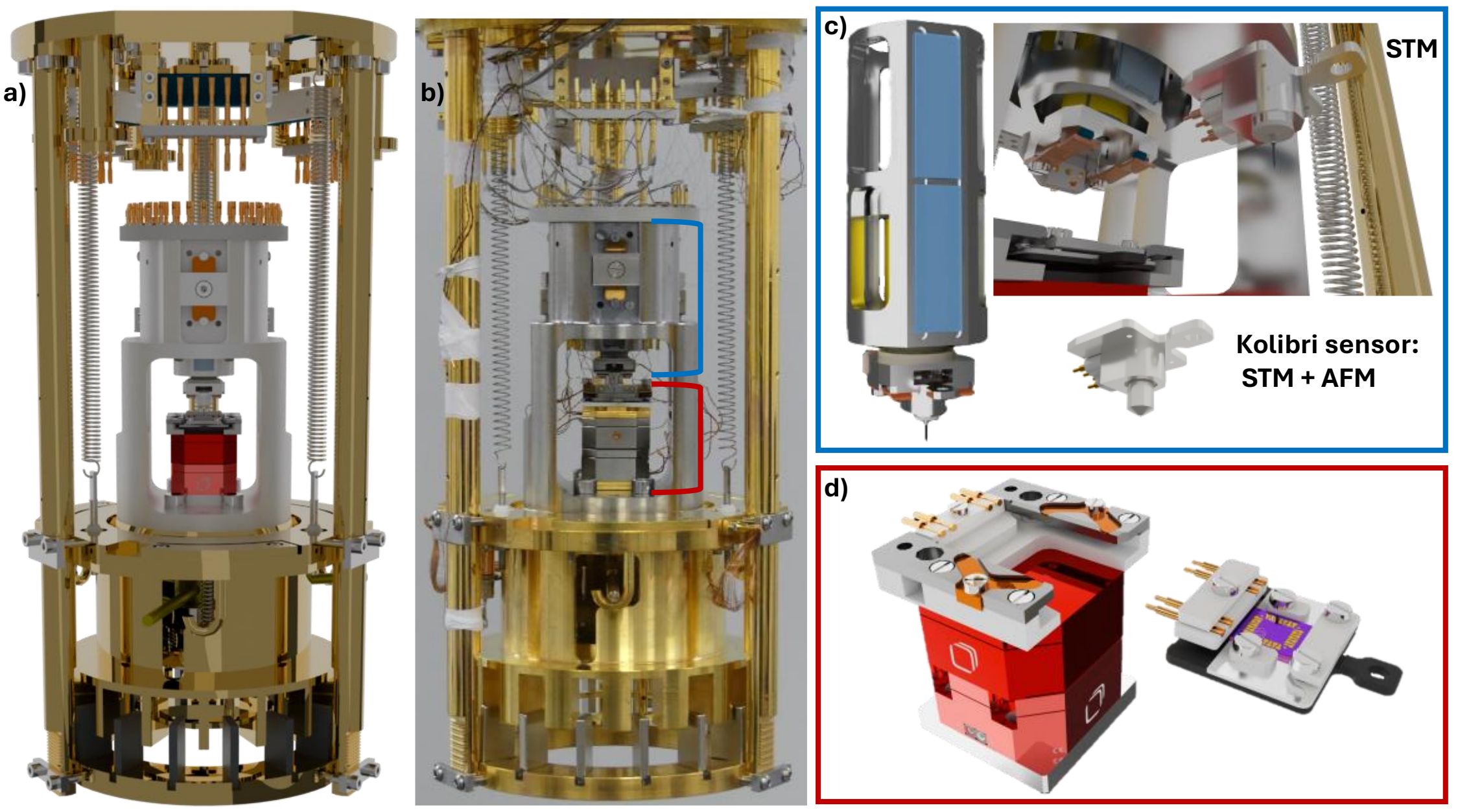

**FIG. 4.** Microscope mounted on gold-plated OFHC copper base. a) 3D design. b) Picture of the actual microscope. c) 3D design of the tip stage showing a STM tip exchanging and a mounted KolibriSensor below. d) 3D design of the sample stage showing the sample exchanging.

## V. Landing on a flake. A drop in an ocean

A key feature of this microscope is its ability to locate and land on specific regions of a sample with 5 µm precision within 10 minutes. This performance relies on two factors: detailed ex situ optical mapping of the 2D device (Fig. 5a–c), and optimized optical access coupled to an optical telescope [29–31], mimicking the setup used in Prof. Crommie's group [32], with enhanced resolution provided by a ×2 amplifier [33]. Together, these elements facilitate a direct line of sight to the sample inside the UHV chamber at a 25° viewing angle (Fig. 5e–g).

The ex situ characterization consists of optically mapping the device and its surroundings, using the gold contacts and local surface inhomogeneities as landmarks to accurately locate the region of interest. For the location of the region of interest, these images are scaled and rotated to match the viewing perspective of the optical telescope, enabling real-time triangulation of the device position (Fig. 5d,g). Once the target region is identified, the tip is approached until its reflection becomes visible. The midpoint between the tip and its reflection defines the landing site with high accuracy (Fig. 5-1).

The final approach proceeds in two steps: first, coarse XY inertial positioning is used to center the 2D device under the tip landing position (Fig. 5, 1–4); second, fine positioning and approach are carried out until tunneling regime is reached (Fig. 5, 5–6). This systematic procedure enables rapid and reliable targeting and characterization of any given sample.

The achieved optical resolution allows us to directly observe the piezotube displacement during scanning and the fine approach into tunneling regime. Targeted landing via optical access works with STM tips prepared using standard methods, such as electrochemically etched tungsten tips or scissor-cut Pt–Ir tips.

By combining high-resolution optical access with this newly developed high-precision region-identification method, specific micrometric regions can be reproducibly located and characterized in different types of samples. This includes not only graphene devices, but also selected few-micron regions in extended homogeneous samples.

In the latter case, specifically for CVD graphene grown on Cu(111), we were able to locate and characterize by STM, in our microscope, the exact same few microns regions containing submicrometric unexpected merging lines that had been previously identified and characterized by AFM [34].

For samples requiring finer location (i.e. transport nanowires) the optical access can be complemented by tuning-fork AFM operation and by large-area capacitive mapping of the tip-sample coupling over the whole accessible XY range. The AFM mode is particularly useful when navigating over non-conductive regions such as $SiO_2$ or hBN, where force-feedback navigation enables a controlled approach before STM measurements and helps prevent unwanted tip–sample contact.

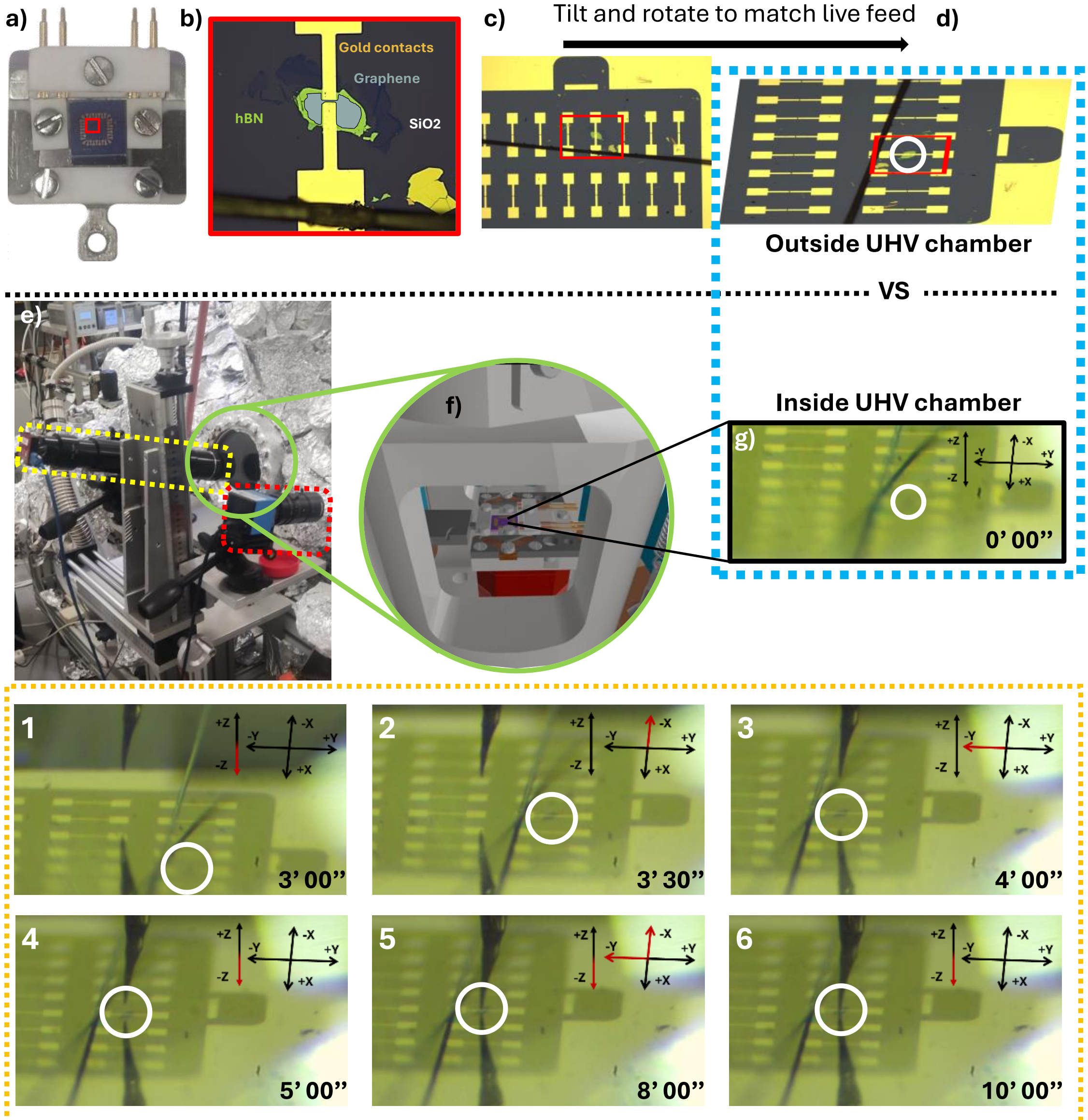

**FIG. 5.** Comparison of external (a-d) and in-situ (e-g) optical characterization, device locating and landing procedure (1-6). Top: a) Image of the sample holder with the mounted sample. b) Optical image of the graphene device with labels. c) External optical microscope image of the device and its surroundings. Same as c) but rotated to fit the angled view obtained with the optical telescope. Middle: e) Optical telescope[29–31,33] attached to the STM chamber (yellow contour) and camera[35] used for sample transfer (red contour). They are mounted on a homemade XY stage to easily target the sample's position. f) 3D representation of the 25º tilted line of sight obtained with the optical telescope with the sample located at the measuring position. g) Actual image of the sample surface inside the UHV chamber, with the targeted region (white circle) out of focus. Bottom: (1–6) Image sequence showing the 10-minute step-by-step localization and landing process on the graphene device. Red x/y (z) arrows indicate the direction of sample (tip) motion at each step. Timer provides a reference for the process duration.

## VI. Performance

To evaluate the performance of the microscope under LT-UHV conditions, we used silicon carbide (SiC) samples[36] and mechanically exfoliated graphene devices fabricated utilizing a custom-build transfer station [8], inspired by the setup employed by Pablo Jarillo-Herrero's group at MIT [9,37,38].

All STM/STS electronics are custom-built. All transport-related measurements were performed using commercial electronics [39–43]. SPM and transport measurements were performed using homemade software WSxM [44]. This in-house infrastructure facilitated the continuous optimization and tailoring of the control systems to the specific operational requirements of the instrument.

## A. Cleanness and mechanical stability

To evaluate the cleanliness of our graphene devices and the vibrational noise levels of the microscope, we measured graphene devices prepared ex-situ. Upon introduction of the samples into the UHV preparation chamber through the load-lock, they were mildly annealed by electron bombardment ~10 h at ~250 ° C. Large-scale STM topography images (Fig. 6, a), revealed completely flat surfaces without polymer residues. Within these regions, smaller-scale STM images routinely yield atomic resolution (Fig. 6, b). Line profiles along these images (Fig. 6, c) reveal a few picometer atomic corrugation, with a vibrational noise below 1 pm.

Vibration transmission through the wiring is minimized by mechanically anchoring each of the electrical lines at several stages. The final section, between the fixed copper base and the spring-suspended microscope head, consists of 40-μm-diameter Kapton-insulated copper wires, which provide a highly flexible connection to the microscope. The four transport lines represent only a small fraction of the full wiring harness, and no degradation of the mechanical stability associated with them has been observed.

Inductive coupling between lines is reduced by using twisted pairs for differential signals up to the final flexible wiring section, by keeping the individual wires separated in the last section, and by grounding the metallic components of the microscope. In particular, the tip holder and surrounding metallic parts are connected to ground, providing electrostatic shielding around the most sensitive STM/AFM signal paths.

The mechanical stability was found to be independent of the sample configuration. Comparable STM resolution and mechanical stability were obtained for STM-only samples and for samples with transport contacts, with the contacts either grounded, connected in the transport configuration, or used to pass a current through the sample. No significant change in the STM topographic stability was observed while passing currents of up to 100 nA through the sample during simultaneous STM imaging.

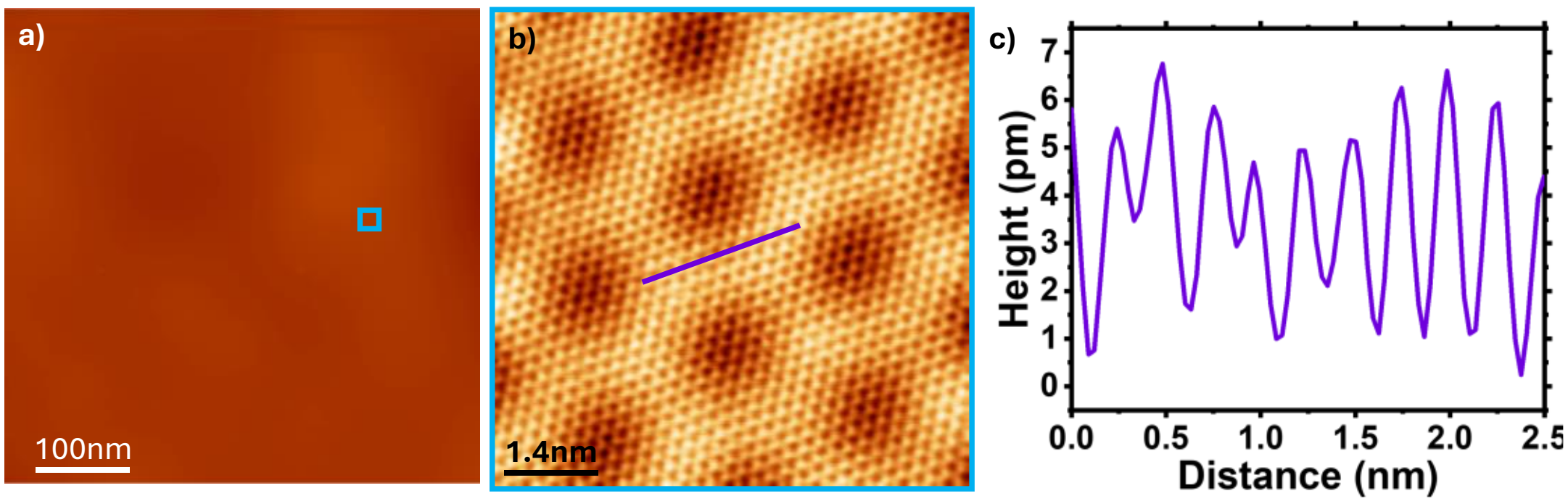

**FIG. 6.** System performance for cleanness and vibrational noise on graphene devices at 4.2K under UHV conditions. a) Large-scale topography image showing a flat featureless surface [I=0.05nA; $V_b$=770mV]. b) Atomic-resolution topography image at the blue square position in a) [I=0.4nA; $V_b$=364mV]. c) Line profile acquired along the purple line in b) showing a vibrational noise below the corrugation of around 1.5 pm.

## B. UHV preparation techniques applied to ultra-clean Gr/hBN devices

The electronic properties of atomically clean 2D graphene devices (like the one shown in Fig. 7) can be tuned by functionalizing their surfaces using a whole set of UHV preparation techniques; approach that, in our opinion, has been underestimated to date. We showed this approach by depositing single atomic hydrogen atoms to introduce local magnetic moments [19] and lead islands to introduce superconductivity by proximity [18] . In both cases, the process started by applying the cleaning procedure described in the previous section, followed by a detailed surface characterization to confirm an atomically clean sample.

Hydrogen deposition was performed using our home-made H-cracker, which induces the thermal dissociation of a beam of $H_2$ passing through a W filament held at 1900K [19]. Subsequent characterization of the surface revealed the detection of single H atoms, surrounded by the characteristic $\sqrt{3}\, x\, \sqrt{3}$ pattern associated with the creation of Kekulé vortices around them[45] (Fig. 7a). We confirmed the possibility of manipulating these H atoms with the STM tip (picking them up and depositing them down at a specific location by applying voltage pulses) [19].

Pb was deposited on graphene devices using a home-made electron-beam metal evaporator. On this graphene surface, we deposit 1 monolayer (ML) of Pb at a rate of 0.5 ML $min^{-1}$ while maintaining the sample at room temperature. As a result, several triangular Pb islands with heights between 2 and 5 nm and sides between 20 and 300 nm were formed, see Fig. 7b.

Altogether, these results validate the system capability to apply UHV preparation and functionalization techniques to atomically clean graphene devices fabricated ex-situ. This methodology will enable the selective incorporation of new properties to 2D devices formed by stacking different vdW materials. Here, we demonstrated the possibility to incorporate magnetism and SC to 2D graphene devices by the controlled absorption of H atoms and Pb islands respectively.

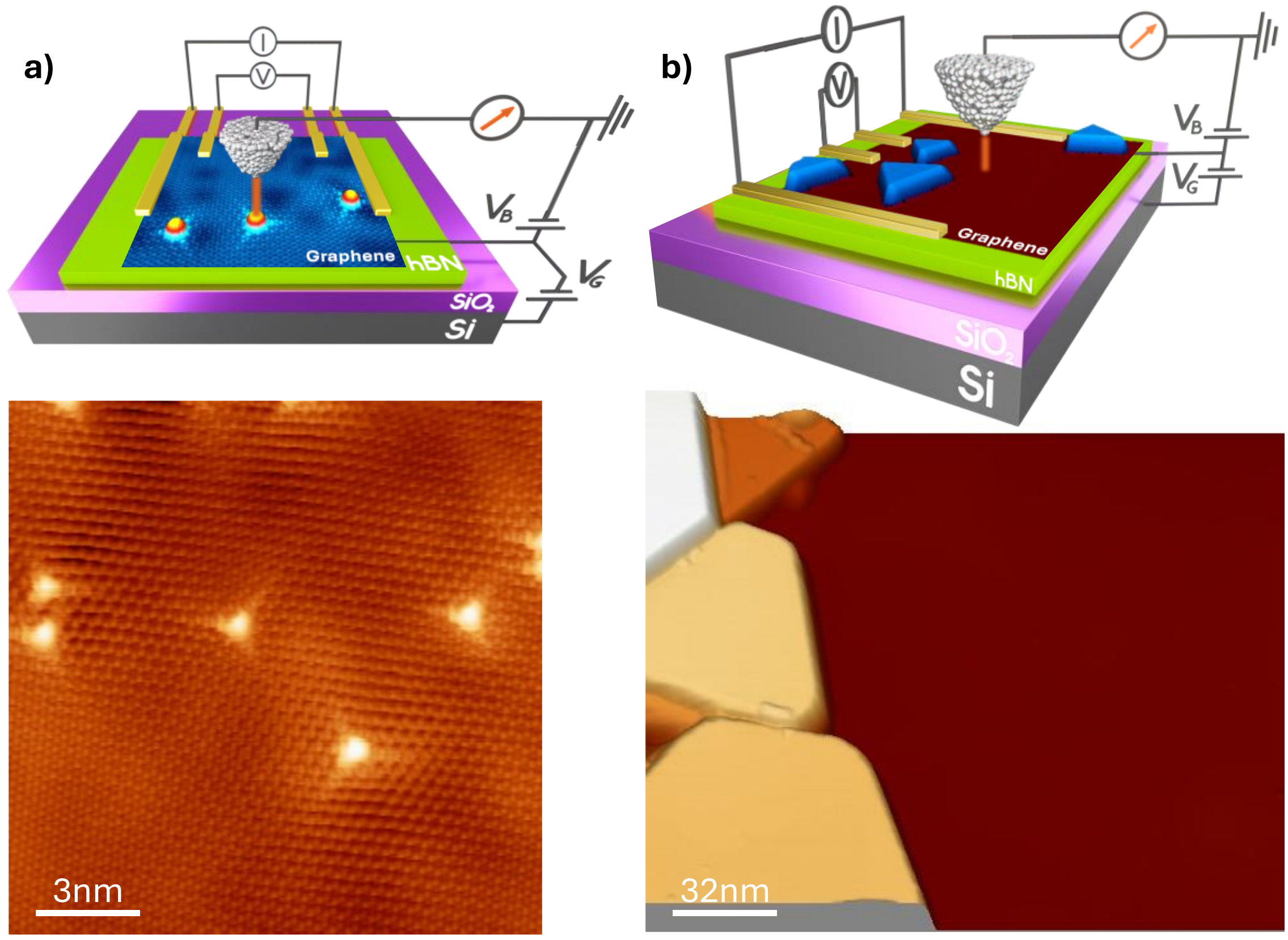

**FIG. 7.** Adding new properties to 2D graphene devices by applying UHV standard sample preparation techniques. a) 3D model (top) and actual topographic image (bottom [I=0.05nA; $V_b$=200mV]) of atomic H atoms deposited on an ultra-clean Gr/hBN/$SiO_2$/Si device using our homemade H cracker. b) 3D model (top) and actual topographic image (bottom [I=0.05nA; $V_b$=1V]) of lead nanoislands deposited using our homemade metal evaporator on a graphene device.

## C. STS Performance

To characterize the STM energy resolution and, in particular, the electronic-noise contribution to the STS energy broadening, we evaporated 300nm of Pb on a graphene sample grown on a SiC(000-1) substrate, which results in a clean, homogeneous lead surface, that we know possess the superconducting properties corresponding to bulk Pb (Fig. 8a)[18]. For these measurements, the cryostat lHe bath was pumped to a base temperature (both probe and sample) of 2.9 K, which is well below the superconducting temperature transition of Pb (Tc = 7.2 K). To increase the energy resolution beyond the thermal limit, and to find our intrinsic electronic noise, we acquire STS spectra using SC Pb tips [18], that we prepared by controlled tip indentations on the Pb surface. The superconducting tunneling gap was characterized by measuring I/V curves on the Pb surface. The differential conductance spectra (*dI/dV vs V*) were obtained by the numerical differentiation of the (*I vs V*) curve. A stabilization current of 0.5 nA and an initial sample voltage of 6.0 mV were used to measure the *I-V* spectrum. By fitting the experimental spectra (black dots in Fig. 8b) to the DOS ansatz proposed by Dynes and coworkers (red line in Fig. 8b) [46], we obtained an upper bound of 30 μV for the voltage-noise-induced energy broadening of the setup.

We emphasize that this value should not be interpreted as a general intrinsic spectroscopic resolution valid for any tip–sample configuration. Rather, it quantifies the contribution of the experimental setup to the energy broadening under highly sensitive superconducting-tip conditions. When normal-metal tips are used, or when the system is operated above the critical temperature of Pb, the effective spectroscopic resolution is mainly determined by this setup-induced broadening together with thermal broadening, and, when applicable, any additional broadening due to the bias modulation.

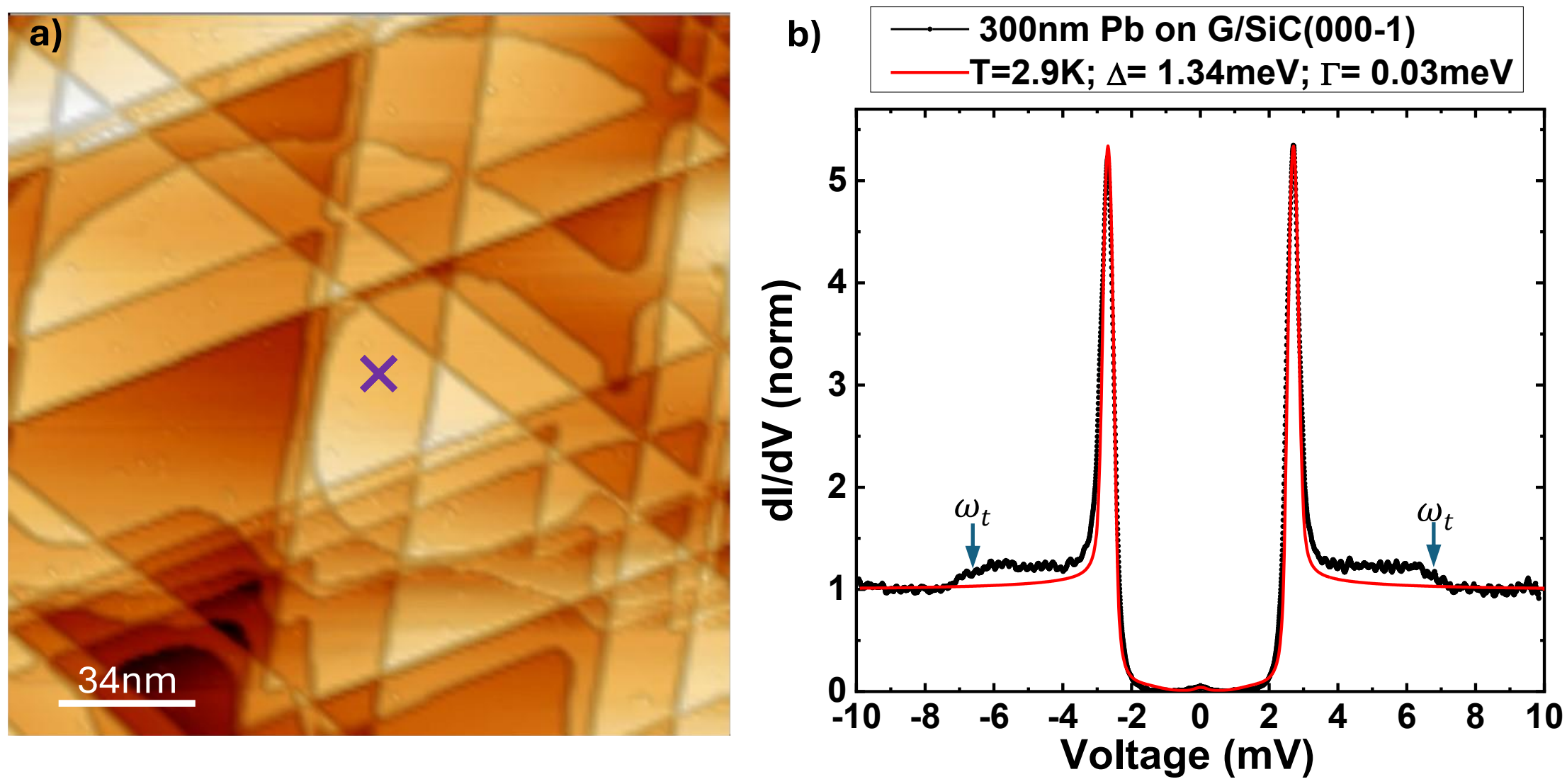

**FIG. 8.** STS System performance. a) Topographic image of the sample's surface, consisting of 300nm of Pb evaporated on a previously annealed G/SiC(000-1) [I=0.05nA; $V_b$=1V]. b) dI/dV curve acquired with a SC Pb tip at the position indicated by the purple cross in a). The experimental data (black) is shown alongside the numerical fit (red) performed using the DOS ansatz proposed by Dynes and coworkers[46]. The comparison between the experimental results and the model yields an energy resolution of 30 μV. Vertical arrows point to the transversal phonons at $V_b=2\Delta+\omega_t\sim6.7$ mV.

## D. Gating Capabilities

One of the primary advantages of measuring micrometer-sized graphene devices is the ability to selectively tune the sample's charge-carrier density (doping). In our setup, the devices use a back-gate geometry in which $V_g$ is applied either to a graphite finger (for non-doped substrates) or directly to the doped silicon substrate (Fig. 9a). Positive (negative) gate voltages induce electron (hole) doping. In graphene, this external electrostatic doping shifts the Dirac point ($E_D$) relative to the Fermi level ($E_F$) by electron or hole filling of the Dirac cone.

To validate the fine tuning of the doping level via the gate voltage, we measured I–V curves at various $V_g$. At $V_g = 0$V, the sample is n-doped due to residual trapped charges, with a charge-neutrality point at $V_{\rm CNP} = -8$V. This doping is enhanced under positive gate voltages (red curve) and shifts toward the p-type regime under negative gate voltages (blue curve) (Fig. 9b). To quantify the gate-induced carrier density, we model the device as a parallel-plate capacitor, with the doped silicon and graphene as electrodes and $SiO_2$ and hBN as the intervening dielectrics. This yields a 2D charge-carrier density $n = \alpha * (V_g - V_{CNP})$ with $\alpha \cong 7.1 \cdot 10^{10} cm^{-2}/V$. In the model, we assume a 285 nm $SiO_2$ layer with $\varepsilon_r = 3.9$ and a 15 nm hBN layer with $\varepsilon_r \approx 3$–4 [47]. Using the relation $E_F = sgn(n)\hbar * v_F * \sqrt{\pi|n|}$ we calculate the expected shift of $E_D$ as a function of $V_g$, accounting for the initial intrinsic doping of the

sample (Fig. 9c, red line). By measuring I–V curves at different $V_g$, we experimentally verified this gating effect by tracking the shift of $E_D$ extracted from the spectra at each gate voltage (Fig. 9c, black dots). The experimental results are in excellent agreement with the calculated model.

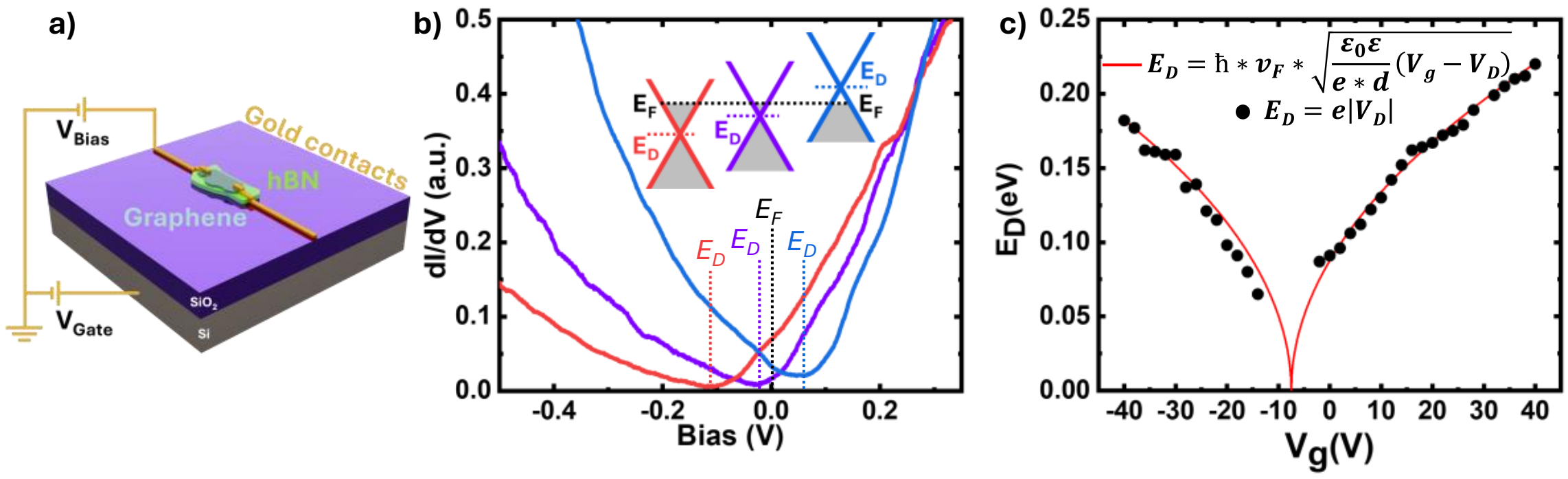

**FIG. 9.** Gating capabilities. a) 3D drawing of a graphene device consisting of a monolayer graphene (MLG) on hexagonal boron nitride (hBN) supported by a doped $SiO_2$/Si substrate. b) IV curves measured on a graphene device at three different gate voltages. c) Evolution of the Dirac Energy ($E_D$) as a function of gate voltage ($V_g$). The red line represents the calculated behavior assuming a parallel-plate capacitor model, while the black dots represent the experimental measurements.

## E. Transport measurements

A standout capability of this microscope is the integration of simultaneous electronic transport measurements. To evaluate the transport performance, we measured the superconducting transition of a custom-fabricated Pb nanowire (Fig. 10). We employed a four-terminal Hall-bar geometry to effectively suppress contact-resistance contributions. The sample was fabricated by evaporating lead within the UHV preparation chamber into a laser-defined photoresist line-pattern on a $SiO_2$/Si chip with pre-evaporated gold contacts. Measurements were carried out in a standard four-probe configuration by sourcing current between the two outermost contacts and recording the resulting voltage drop between the inner contacts (Fig. 10a). We first measured *V/I* curves at I=100μA for sample temperatures ($T_s$) between 4.5 and 20.5K (Fig. 10b). For temperatures above ~7.3K, V/I exhibit linear Ohmic behavior, with the resistance (determined by the slope) decreasing with the temperature. For temperatures below ~7.3K, the voltage drop vanishes, signaling the onset of the superconducting regime. The temperature-dependent resistance extracted from these slopes shows a sharp drop to zero at the critical temperature, marking the transition from a normal metal to a superconductor (Fig. 10e, green dashed line).

An alternative method to characterize the superconducting transition is through the measurement of the critical current, $I_c$, defined as the maximum bias current that can be applied without the onset of a measurable voltage. In quasi-one-dimensional Pb nanowires, the observed $I_c$ is often governed by phase-slip processes and/or thermal instabilities, while the depairing (pair-breaking) current —representing the intrinsic maximum supercurrent of the wire— sets an upper bound. To determine this threshold, we measured *V/I* characteristics at progressively higher bias currents over a range of temperatures (Fig. 10d). For $I < I_c$, the voltage remains below the measurement resolution, producing a near-zero-voltage plateau. As the current approaches $I_c$, the curves switch from this plateau to a finite-slope regime, consistent with the onset of dissipation and a nonzero differential resistance. From these data we extract $I_c(T)$ (Fig. 10e), which decreases with increasing temperature and vanishes at T ≈ 7.3 K, identifying the superconducting-to-normal transition, (Fig. 10e, black dashed line).

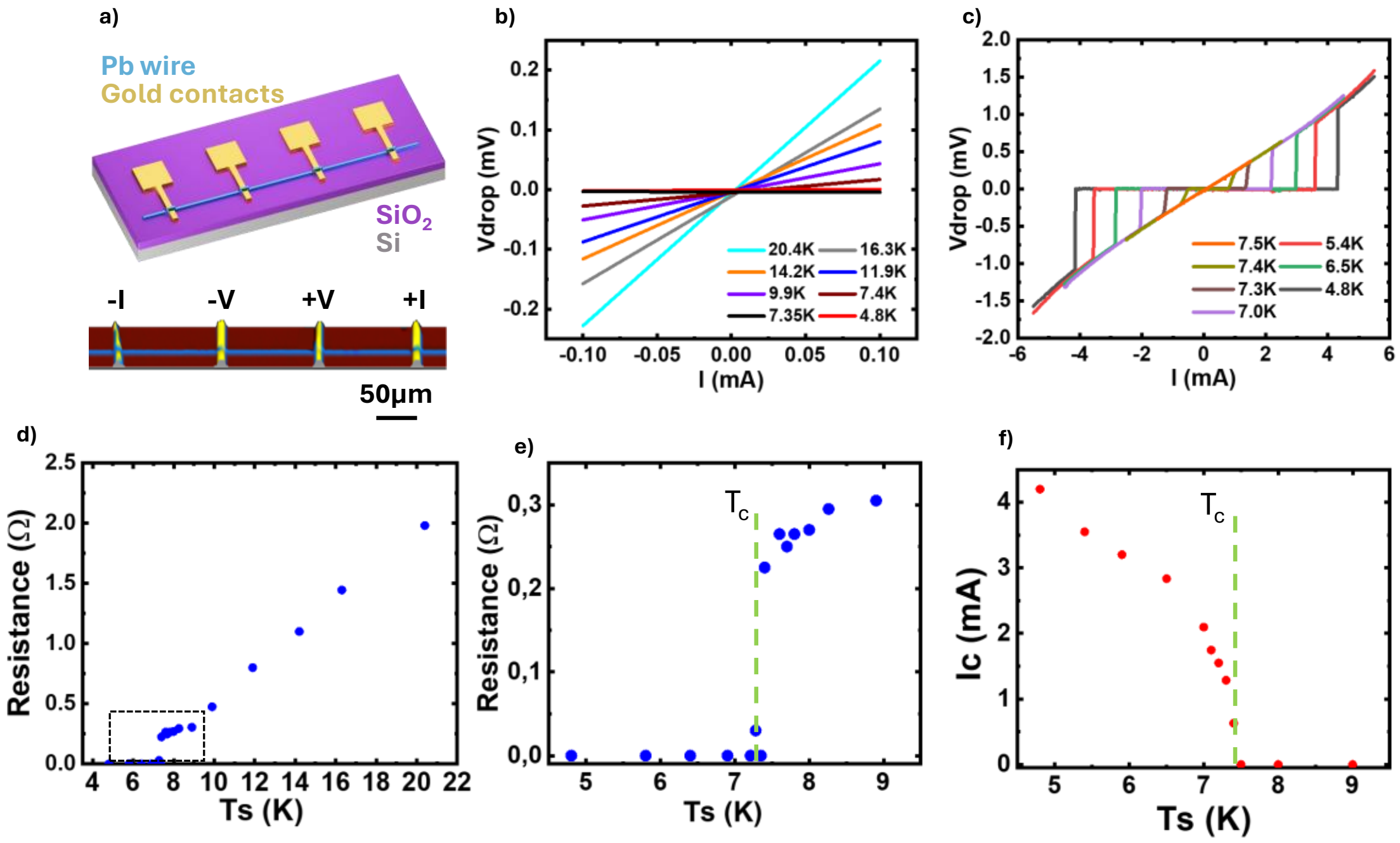

**FIG. 10.** Measuring the superconducting transition of a Pb nanowire. a) From top to bottom: 3D design of the fabricated nanowire and optical microscope image processed for 3D visualization. The four-terminal configuration for the measurements is indicated. b) Four-terminal VI curves measured at different temperatures in the low current regime. c) Four-terminal VI curves acquired at different temperatures in the high current regime. d) Temperature dependence of resistance obtained from b). e) Zoom in corresponding to the dashed square outlined in d), showing the temperature dependence of the resistance across $T_c$. f) Critical current ($I_c$) temperature dependance obtained from c).

## VII. Summary and conclusion

We have presented the development of a new LT-UHV-STM/AFM capable of landing on a μm-sized sample with 5μm precision in less than 10 minutes. We have demonstrated the STM performance on both bulk-like Pb samples and graphene devices, highlighting an energy resolution of 30μV and a vibrational noise below 1pm. Furthermore, we have shown that our full machinery of in situ UHV sample preparation techniques can be effectively applied to ultra-clean graphene devices fabricated by ourselves, where electronic doping has been tuned using a back gate voltage. Finally, transport capabilities where demonstrated by measuring the superconducting transition of an in-situ fabricated Pb nanowire, observing the characteristic drops in both electrical resistance and critical current at the SC critical temperature.

## VIII. Acknowledgments

We thank Wolfgang Stiepany and Klaus Kern for their key guidance during the first stages of the design. We thank Eduardo Lee and his group for all the asistance implementing the transport setup and for many useful discussions. We thank Guillermo López-Polín, Lucía Romero-Sánchez and Héctor-González-Herrero for all the samples provided to test the performance of the microscope. We also thank Lucía Romero-Sánchez for contributing on the Pb-wire transport measurements. We acknowledge financial support from the Spanish Ministry of Science and Innovation, through project PID2023-149106NB-I00, the Comunidad de Madrid and the Spanish State through the Recovery, Transformation and Resilience Plan [Materiales Disruptivos Bidimensionales (2D), (MAD2DCM)-UAM Materiales Avanzados], the European Union through the Next Generation EU funds to the project "Artificial Quantum Matter: From 2D Materials to Spin Lattice Systems"from the BBVA Foundation Fundamentos Program 2024.

## IX. Author Declarations

### A. Conflict of Interest

The authors have no conflicts to disclose.

### B. Author Contributions

Diego Exposito: Conceptualization (equal); Design (lead); Construction (lead); Measurements (equal), Writing (lead).

Oscar Custance: Design (supporting); Writing (supporting)

Iván Brihuega: Conceptualization (equal); Measurements (equal); Supervision (lead); Funding acquisition (lead); Writing (supporting)

## X. Data availability

The data that support the findings of this study are available from the corresponding author upon reasonable request.